
\magnification=\magstep 2
\overfullrule=0pt
\hfuzz=16pt
\voffset=0.0 true in
\vsize=8.8 true in
\baselineskip 20pt
\parskip 6pt
\hoffset=0.1 true in
\hsize=6.3 true in
\nopagenumbers
\pageno=1
\footline={\hfil -- {\folio} -- \hfil}

\ 

\

\centerline{\bf Recent Theoretical Results for Nonequilibrium}

\centerline{\bf Deposition of Submicron Particles}

\vskip 0.4in

\centerline{\bf Vladimir Privman}

\vskip 0.2in

\centerline{\sl Department of Physics, Clarkson University,
Potsdam, New York 13699--5820, USA}

\vskip 0.4in

\centerline{\bf ABSTRACT}

Selected theoretical
developments in modeling of deposition of submicrometer size 
(submicron) particles on
solid surfaces, with
and without surface diffusion, of interest in colloid, polymer, and 
certain biological systems, are surveyed.
We review deposition processes involving
extended objects, with jamming and its interplay with
in-surface diffusion yielding interesting dynamics of
approach to the large-time state. 
Mean-field and low-density 
approximation schemes can be used in many instances 
for short and intermediate times, in large enough dimensions, 
and for particle 
sizes larger than few
lattice units. Random sequential adsorption models are appropriate
for higher particle densities (larger times). Added diffusion 
allows formation of denser
deposits and leads to power-law large-time behavior which,
in one dimension (linear substrate, such as DNA), was related 
to diffusion-limited reactions, while
in two dimensions (planar substrate), was associated with evolution 
of the domain-wall and defect network, reminiscent of equilibrium ordering
processes.

\ 

\noindent\hrule 

\ 

\noindent{}This is a review article, to appear in {\it The Journal of Adhesion\/} (2000).

\vfill


\vfil\eject

\noindent{\bf 1. Introduction}

\noindent{\it 1.1. Surface Deposition of Submicron Particles}

Surface deposition of submicron particles is of immense practical importance [1-4]. Typically, particles of this size, colloid, protein or other biological objects, are suspended in solution, without sedimentation due to gravity. In order to maintain the suspension stable, one has to prevent aggregation (coagulation) that results in larger flocks for which gravity pull is more profound. Stabilization by particle-particle electrostatic repulsion or by steric effects, etc., is usually effective for a sufficiently dilute suspension. But this means that even if a well-defined
suspension of well-characterized particles is available, it cannot be always
easily observed experimentally in the bulk for a wide range of particle interactions. For those interaction parameters for which the system is unstable with respect to coagulation, the time of observation will be
limited by the coagulation process which can be quite fast.

One can form a dense deposit slowly, if desired, {\it on a surface}. Indeed,
particles can be deposited by diffusion, or more realistically by convective diffusion [5] from a flowing suspension, on collector surfaces. The suspension itself need not be dense even though the on-surface deposit might
be quite dense, depending of the particle-particle and particle-surface
interactions. Dilution of suspension generally prolongs an experiment aimed at reaching a certain surface coverage. Thus, surface deposition has been well established as an important tool to probe interactions of matter
objects on the submicron scale [1-4].

\noindent{\it 1.2. Particle Jamming and Screening at Surfaces}

Figure 1 illustrates possible configurations of particles at a surface. From left to right, we show particles
deposited on the surface of a collector, then particles deposited on top of other particles. The latter
is possible only in the absence of significant particle-particle repulsion. The two situations are termed monolayer and multilayer deposition even though the notion of a layer beyond the one exactly at the surface is only approximate. We next show two effects that play important role in surface growth. The first is jamming: a particle marked by an open circle cannot fit in the lowest layer at the surface. 
A more realistic two-dimensional ($2D$) configuration is shown in the inset. 

The second effect is screening:
surface position marked by the open circle is not reachable. Typically, in colloid deposition monolayer or few-layer deposits are formed and the dominant effect is jamming, as will be discussed later. Screening plays dominant role in deposition of multiple layers and, together with the transport mechanism, determines the morphology of the growing surface. In addition, the configuration on the surface depends on the transport mechanism of the particles to it and on the particle motion on the surface, as well as possible
detachment. Particle motion is typically
negligible for colloidal particles but may be significant for proteins.

\noindent{\it 1.3. Role of Dimensionality and Relation to Other Systems}

An important feature of surface deposition is that for all practical purposes it is essentially a $2D$
problem. As a result, any mean-field, rate-equation, effective-field, etc., approaches which are
usually all related in that they ignore long-range correlations and fluctuation effects, 
may not be applicable.
Indeed, it is known that as the dimensionality of a many-body interacting system 
decreases, fluctuations
play a larger role. 
Dynamics of important physical, chemical, and biological
processes [6-7] provides examples of strongly fluctuating systems in low
dimensions, $D=1$ or 2. These processes include surface
adsorption on planar substrates or on large collectors.
The surface of the latter is semi-two-dimensional owing
to their large size as compared to the size of the deposited 
particles.
 
The classical chemical reaction-diffusion
kinetics corresponds to
$D=3$. However, heterogeneous catalysis
generated interest in $D=2$. For both deposition
and reactions, some experimental results exist even in
$D=1$ (see 
later). Finally, kinetics of ordering and phase separation,
largely amenable to experimental probe in $D=3$ and $2$, 
attracted much
recent theoretical effort in $D=1,2$.

Models in $D=1$, and sometimes 
in $D=2$, allow
derivation of analytical results. Furthermore, it turns out
that all three types of model: deposition-relaxation,
reaction-diffusion, phase separation, are interrelated in
many, but not all, of their properties. This observation is
by no means obvious. It is model-dependent and
can be firmly established [6-7] only in low
dimensions, mostly in $D=1$.

Such low-dimensional nonequilibrium models pose
several interesting challenges theoretically and
numerically. While many exact, asymptotic, and numerical
results are already available in the literature [6-7], this
field presently provides examples of properties 
which lack theoretical explanation
even in $1D$. Numerical simulations are challenging and
require large scale computational effort already for $1D$
models. For more experimentally relevant $2D$ cases, where
analytical results are scarce, difficulty in numerical
simulations has been the limiting factor in
understanding of many open problems.

\noindent{\it 1.4. Outline of This Review}

The purpose of this article is to provide an introduction to
the field of nonequilibrium surface deposition models of
extended particles. By ``extended'' we mean that the main
particle-particle interaction effect will be jamming, i.e., mutual exclusion.
No comprehensive survey of the
literature is attempted. Relation of deposition to other
low-dimensional models mentioned earlier will be only
referred to in detail in few cases. The specific models
and examples selected for a more detailed exposition, i.e.,
models of deposition with diffusional relaxation, were
biased by author's own work.

The outline of the review is as follows. The rest of this
introductory section is devoted to defining the specific
topics of surface deposition to be surveyed. Section~2
describes the simplest models of random sequential
adsorption. Section~3 is devoted to deposition with
relaxation, with general remarks followed by definition of
the simplest, $1D$ models of diffusional relaxation for
which we present a more detailed description of various
theoretical results. Multilayer deposition is also
commented on in Section~3. More numerically-based $2D$ results
for deposition with diffusional relaxation are surveyed in
Section~4. Section~5 presents brief concluding remarks.

Surface deposition is a vast field of study. 
Our emphasis here will be on
those deposition processes where the particles are
``large'' as compared to the underlying atomic and
morphological structure of the substrate and as compared to
the range of the particle-particle and particle-substrate
interactions. Thus, colloids, for instance, involve particles
of submicron to several micron size. 
We note that 1$\mu$m$=
10000$\AA, whereas atomic dimensions are of order 1\AA,
while the range over which
particle-surface and particle-particle interactions are significant
as compared to $kT$, is typically of order 100\AA{}
or less.
Extensive theoretical study of such systems
is relatively recent and it has been motivated by
experiments where submicron-size colloid, polymer, and
protein ``particles'' were the deposited objects [1-4,8-18].

Perhaps the simplest and the most studied model with
particle exclusion is Random Sequential Adsorption (RSA). 
The RSA
model, to be described in detail in Section~2, assumes that
particle transport (incoming flux) onto the surface results in
a uniform deposition attempt rate $R$ per unit time and area. 
In the simplest formulation, one assumes that only
monolayer deposition is allowed. Within this
monolayer deposit, each new arriving particle must either
fit in an empty area allowed by the hard-core
exclusion interaction with the particles deposited earlier,
or the deposition attempt is rejected.

The basic RSA model will be described shortly,
in Section~2. Recent work has been focused on its
extensions to allow for particle relaxation by diffusion, 
see Sections~3 and 4, to include detachment processes, and to
allow multilayer formation. The latter two extensions will
be briefly surveyed in Section~3. Several other
extensions will not be discussed [1-4].

\vfil\eject

\noindent{\bf 2. Random Sequential Adsorption}

\noindent{\it 2.1. The RSA Model}

The irreversible Random Sequential Adsorption
(RSA) process [19-20] models
experiments of
submicron particle deposition by assuming
 a planar $2D$ substrate and, in the simplest case,
continuum (off-lattice) deposition of spherical particles.
However, other RSA models have also received attention. In $2D$, 
noncircular cross-section shapes as well as various
lattice-deposition models were considered [19-20].
Several experiments on polymers and
attachment of fluorescent units on DNA
molecules [18] (the latter is usually accompanied
by motion of these units on the DNA and
detachment) suggest consideration of the lattice-substrate
RSA processes in $1D$. RSA processes have also found
applications in traffic problems and certain other fields.
Our presentation in this section aims at
defining some RSA models and outlining characteristic
features of their dynamics.

Figure~2 illustrates the simplest possible monolayer
lattice RSA model: irreversible deposition of dimers on
the linear lattice. An arriving dimer 
will be deposited if the underlying
pair of lattice sites are both empty. Otherwise, it is discarded, which is
shown schematically by the two dimers above the surface layer. 
Their deposition on the surface is not possible unless detachment
and/or motion of monomers or whole dimers clear the appropriate
landing sites. 

Let us consider the irreversible RSA without detachment or diffusion.
The substrate is usually
assumed to be empty initially, at $t=0$. In the course of time $t$, the
coverage, $\rho (t)$, increases and builds up to order 1 on
the time scales of order $\left( R V \right)^{-1}$, where
$R$ was defined earlier as the deposition attempt rate per
unit time and area of the surface,
while $V$ is the particle $D$-dimensional ``volume.''
For deposition of spheres on a planar surface, $V$ is actually
the cross-sectional area.

At large times the coverage approaches the jammed-state
value where only gaps smaller than the particle size were
left in the monolayer. The resulting state is less dense
than the fully ordered close-packed coverage. For the
$D=1$ deposition shown in Figure~2 the fully ordered state
would have $\rho =1$. The variation of the RSA coverage is
illustrated by the lower curve in Figure~3.

At early times the monolayer deposit is not dense and the
deposition events are largely uncorrelated. In this regime,
mean-field like low-density approximation schemes are
useful [21-23]. Deposition of $k$-mer particles on the linear
lattice in $1D$ was in fact solved exactly for all times [24].
In $D=2$, extensive
numerical studies were reported [23,25-36] of the variation of
coverage with time and large-time asymptotic behavior
which will be discussed shortly. Some exact
results [24] for correlation properties are available 
in $1D$. Numerical results [27] for correlation properties have 
been obtained in $2D$.

\noindent{\it 2.2. The Large-Time Behavior in RSA}

The large-time deposit has several characteristic properties. 
For lattice
models, the approach to the jammed-state coverage is
exponential [36-38]. This was shown to follow from the
property that the final stages of deposition are in few
sparse, well separated surviving landing sites.
Estimates of decrease in their density at late stages
suggest that

$$ \rho(\infty ) - \rho(t) \sim \exp \left( -R \ell^D t
\right) \;\; , \eqno(1) $$

\noindent{}where $\ell$ is the lattice spacing and $D$ is
the dimensionality of the substrate. The
coefficient in Eq.~(1) is of order $\ell^D/V$ if the coverage
is defined as the fraction of lattice units covered, i.e.,
the dimensionless fraction of area covered, also termed the
coverage fraction, so that coverage as density of particles
per unit volume would be $V^{-1} \rho$. The detailed
behavior depends of the size and shape of the depositing
particles as compared to the underlying lattice unit
cells.

However, for continuum off-lattice deposition, formally
obtained as the limit $\ell \to 0$, the approach to the
jamming coverage is power-law. This interesting behavior [37-38]
is due to the fact that for large times the remaining
voids accessible to particle deposition can be of sizes
arbitrarily close to those of the depositing particles.
Such voids are thus reached with very low probability by the
depositing particles, the flux of which is uniformly
distributed. The resulting power-law behavior depends on
the dimensionality and particle shape. For instance, for
$D$-dimensional cubes of volume $V$,

$$ \rho(\infty ) - \rho(t) \sim { \left[\ln ( RVt )
\right]^{D-1} \over RVt } \;\; , \eqno(2) $$

\noindent{}while for spherical particles,

$$ \rho(\infty ) - \rho(t) \sim ( RVt )^{-1/D} \;\; .
\eqno(3) $$

For $D>1$, the expressions Eqs.~(2-3), and similar relations
for other particle shapes, are actually empirical
asymptotic laws which have been verified, mostly for $D=2$,
by extensive numerical simulations [4,25-36].
The most studied $2D$ geometries are circles
(corresponding to the deposition of spheres on a
planar substrate) and squares. The jamming coverages are

$$ \rho_{\rm squares}
(\infty) \simeq 0.5620 \;\;\;\;\; {\rm and}
\;\;\;\;\; \rho_{\rm circles}(\infty) \simeq 0.544 
\; {\rm to} \; 0.550 \;\; , \eqno(4) $$

\noindent{}much lower than the close-packing values, 
1 and ${\pi \over 2 \sqrt{3}} \simeq 0.907$, respectively.
For square particles, the crossover to continuum
in the limit $k \to \infty$ and $\ell \to 0$, with fixed
$V^{1/D}=k\ell$ in deposition of $k \times k \times \ldots
\times k$ lattice squares, has been investigated in some
detail [36], both analytically (in any $D$) and numerically
(in $2D$).

The correlations in the large-time jammed state are
different from those of the equilibrium random gas of
particles with density near $\rho (\infty )$. In fact, the
two-particle correlations in continuum deposition develop a
weak singularity at contact, and correlations generally
reflect the full irreversibility of the
RSA process [24,27,38]. 

\vfil\eject

\noindent{\bf 3.~Deposition with Relaxation}

\noindent{\it 3.1. Detachment and Diffusional Relaxation}

Monolayer deposits may relax, i.e., explore more
configurations, by particle motion on the
surface, by their detachment, as well as by motion
and detachment of the constituent monomers or
recombined units. In fact,
detachment has been
experimentally observed in deposition of colloid particles
which were otherwise quite immobile on the surface [39].
Theoretical interpretation of colloid particle detachment data
has proved difficult, however, because binding to the substrate
once deposited, can be different for different particles, whereas
the transport to the substrate, i.e., the flux of
the arriving particles in the deposition part of the
process, typically by convective diffusion, is more
uniform. Detachment also plays role in deposition on DNA
molecules [18].

Recently, more theoretically motivated studies of the
detachment relaxation processes, in some instances with
surface diffusion allowed as well, have lead to interesting
model studies [40-46]. These investigations did not always assume
detachment of the original units. Models involving monomer
recombination prior to detachment, of $k$-mers in
$D=1$, have been mapped onto certain spin models and symmetry
relations were identified which allowed derivation of several
exact and asymptotic results on the
correlations and other properties [40-46]. 
We note that deposition and detachment combine to
drive the dynamics into a steady state, rather than jammed
state as in ordinary RSA. These studies 
have been largely limited thus far to $1D$ models.

We now turn to particle motion on the surface, in a
monolayer deposit, which was experimentally observed in
deposition of proteins [17] and also in deposition on DNA
molecules [18,47]. 
From now on, we consider diffusional
relaxation, i.e., random hopping on the surface in the lattice case. 
The dimer deposition in $1D$, for instance, is shown in Figure~2.
Hopping of dimer particles one site to the left or to the
right is allowed only if the target site is not occupied.
Such hopping can open a two-site gap to allow additional 
deposition. Thus, diffusional relaxation lets the
deposition process to reach denser, in fact, close-packed
configurations. Initially, for short times, when the empty area is
plentiful, the effect of the in-surface particle motion
will be small. However, for large times, the density will
exceed that of the RSA process, as illustrated by the
upper curve in Figure~3.

It is important to emphasize that deposition and diffusion
are two independent processes going on at the same time.
External particles arrive at the surface with a fixed rate
per unit area. Those finding open landing sites are
deposited; others are discarded. At the same time, internal 
particles, those already on the surface, attempt, with some
rate, to hop to a nearby site. They actually move only 
if the target site is available. 

\noindent{\it 3.2. One-Dimensional Models}

Further investigation of this effect is much simpler in
$1D$ than in $2D$. Let
us therefore consider the $1D$ case first, postponing the
discussion of $2D$ models to the next section.
Specifically, consider deposition of $k$-mers of fixed
length $V$. By keeping the length fixed, we can also naturally
consider the continuum limit of no lattice by having the
lattice spacing vanish as $k \to \infty$. This limit
corresponds to continuum deposition if we take the
underlying lattice spacing $ \ell = V/k$. Since the
deposition attempt rate $R$ was defined per unit area (unit
length here), it has no significant $k$-dependence. However,
the added diffusional hopping of $k$-mers on the $1D$
lattice, with the attempt rate to be denoted by $H$, 
and hard-core or similar
particle interaction, must be $k$-dependent. Indeed, we
consider each deposited $k$-mer particle as randomly and
independently attempting to move one lattice spacing to the
left or to the right with the rate $H/2$ per unit time. 
Particles cannot run over each other so some sort
of hard-core interaction must be assumed, i.e., in a
dense state most hopping attempts will fail. However, if
left alone, each particle would move diffusively for large
time scales. In order to have the resulting diffusion
constant $\cal D$ finite in the continuum limit $k \to
\infty$, we must assume that

$$ H \propto {\cal D} / \ell^2 = {\cal D}k^2 /
V^2\;\; . \eqno(5) $$

\noindent{}which is only valid in $1D$.

Each successful hopping of a particle results in motion of
one empty lattice site. It
is useful to reconsider the dynamics of particle hopping in
terms of the dynamics of this rearrangement of empty area
fragments [48-50]. Indeed, if several of these empty sites are
combined to form large enough voids, deposition attempts
can succeed in regions of particle density which would be
jammed in the ordinary RSA. In terms of these new
``diffuser particles'' which are the empty lattice sites of the
deposition problem, the process is in fact that of
reaction-diffusion. Indeed, $k$ reactants (empty sites)
must be brought together by diffusional hopping in order
to have finite probability of their annihilation, i.e.,
disappearance of a group of consecutive nearest-neighbor
empty sites due to successful deposition. Of course, the
$k$-group can also be broken apart due to diffusion.
Therefore, the $k$-reactant annihilation is not
instantaneous in the reaction nomenclature. Such
$k$-particle reactions are of interest on their own [51-57].

\noindent{\it 3.3. Beyond the Mean-Field Approximation}

The simplest mean-field rate equation for annihilation of
$k$ reactants describes the time dependence of the
coverage, $\rho (t)$, in terms of the reactant density
$1-\rho$, i.e., the density of the empty spaces,

$$ {d \rho \over dt} = \Gamma (1-\rho)^k \;\; , \eqno(6)
$$

\noindent{}where $\Gamma$ is the effective rate constant.
Note that we assume that the close-packing dimensional
coverage is 1 in $1D$. There are two problems 
with this approximation. Firstly,
it turns out that for $k=2$ the mean-field approach breaks
down. Diffusive-fluctuation arguments for non-mean-field
behavior have been advanced for several chemical 
reactions [51,53,59-59].
In $1D$, several exact calculations support this
conclusion [60-66]. The asymptotic
large-time behavior turns out to be

$$ 1-\rho \sim 1/\sqrt{t} \;\;\;\;\;\;\;\; (k=2,D=1) \;\; ,
\eqno(7) $$

\noindent{}rather than the mean-field prediction $\sim
1/t$. The coefficient in Eq.~(7) is expected to be universal, 
when expressed in an appropriate dimensionless form by
introducing single-reactant diffusion constant.

The power law Eq.~(7) was confirmed by extensive
numerical simulations of dimer deposition [67] and by exact
solution [68] for one particular value of $H$ for a model
with dimer dissociation. The latter work
also yielded some exact results for correlations.
Specifically, while the connected
particle-particle correlations spread diffusively in space,
their decay it time is nondiffusive [68]. 
Series expansion studies of models of dimer deposition with
diffusional hopping of the whole dimers or their dissociation
into hopping monomers, has confirmed the expected asymptotic behavior
and also provided estimates of the coverage as a function of time [69].

The case $k=3$ is marginal with the mean-field power law
modified by logarithmic terms. The latter were not observed
in Monte Carlo studies of deposition [49]. However,
extensive results are available directly for three-body
reactions [53-56], including verification of the
logarithmic corrections to the mean-field behavior [54-56].

\noindent{\it 3.4. Continuum Limit of Off-Lattice Deposition}

The second problem with the mean-field rate equation is
identified when one attempts to use it in the continuum 
limit corresponding to off-lattice
deposition, i.e., for $k \to \infty$. Note that Eq.~(6) has no regular
limit as $k\to \infty$. The
mean-field approach is essentially the fast diffusion
approximation assuming that diffusional relaxation is
efficient enough to equilibrate nonuniform density
fluctuations on
the time scales fast as compared to the time scales of the
deposition events. Thus, the mean-field results are
formulated in terms of the uniform properties, such as
the density. It turns out, however, that the simplest, $k^{\rm
th}$-power of the reactant density form Eq.~(6) is only
appropriate for times $t >> e^{k-1}/(RV)$.

This conclusion was reached [48] by assuming the
fast-diffusion, randomized hard-core reactant
system form of the inter-reactant distribution function in
$1D$. This approach, not detailed here, allows
estimation of the limits of
validity of the mean-field results and it correctly
suggests mean-field validity for $k=4,5,\ldots$, with
logarithmic corrections for $k=3$ and complete breakdown of
the mean-field assumptions for $k=2$. This
detailed analysis yields the modified mean-field relation

$$ {d \rho \over dt } = {\gamma RV (1-\rho)^k \over \left(
1-\rho + k^{-1}\rho \right) } \;\;\;\;\;\;\;\; (D=1)
\;\; , \eqno(8) $$

\noindent{}where $\gamma$ is some effective dimensionless
rate constant. This new expression applies
uniformly as $k \to \infty$. Thus, the continuum deposition
is also asymptotically mean-field, with the
essentially-singular rate equation

$$ {d \rho \over dt} = \gamma (1-\rho) \exp [-\rho / (1-\rho)]
\;\;\;\;\;\;\;\; (k=\infty,D=1) \;\; . \eqno(9) $$

\noindent{}The approach to the full, saturation coverage
for large times is extremely slow,

$$ 1 - \rho (t) \approx {1 \over \ln \left( t \ln t \right) }
\;\;\;\;\;\;\;\; (k=\infty,D=1) \;\; . \eqno(10) $$

\noindent{}Similar predictions were also derived
for $k$-particle chemical reactions [53].

\noindent{\it 3.5. Comments on Multilayer Deposition}

When particles are allowed to attach 
also on top of each other, with possibly some
rearrangement processes allowed as well, multilayer
deposits will be formed. It is important to note that the
large-layer structure of the deposit and fluctuation
properties of the growing surface will be determined by the
transport mechanism of particles to the surface and
by the allowed relaxations (rearrangements). Indeed, these
two characteristics determine the screening properties of
the multilayer formation process which in turn shape the
deposit morphology, which can range from fractal to dense,
and the roughening of the growing deposit surface. There is a
large body of research studying such growth, with recent
emphasis on the growing surface fluctuation properties.

However, the feature characteristic of the RSA
process, i.e., the exclusion due to particle size, plays no
role in determining the universal, large-scale properties
of thick deposits and their surfaces. Indeed, the
RSA-like jamming will be only important for detailed
morphology of the first few layers in a multilayer
deposit. However, it turns out that RSA-like
approaches (with relaxation) can be useful in
modeling granular compaction [70].

In view of the above remarks, multilayer deposition 
models involving jamming effects were relatively
less studied. They can be divided into two groups. Firstly,
structure of the deposit in the first few layers is of
interest [71-73] because they retain memory of the surface.
Variation of density and other correlation properties away
from the wall has structure on the length scale of
particle size. These typically oscillatory
features decay away with the distance from the wall.
Numerical Monte Carlo simulation aspects
of continuum multilayer
deposition (ballistic deposition of $3D$ balls)
were reviewed in [73].
Secondly, few-layer deposition processes have been of
interest in some experimental systems. Mean-field
theories of multilayer deposition with particle size and
interactions accounted for were formulated [74] and used to fit
such data [15-16,75-76].

\vfil\eject

\noindent{\bf 4. Two-Dimensional Deposition with
Diffusional Relaxation}

\noindent{\it 4.1. Combined Effects of Jamming and Diffusion}

We now turn to the $2D$ case of deposition of extended
objects on planar surfaces, accompanied by diffusional
relaxation, assuming monolayer deposits. We note that the
available theoretical results are limited to few 
studies [34,77-79]. They indicate a rich pattern of new
effects as compared to 
$1D$. In fact, there exists extensive literature [81] on
deposition with diffusional relaxation in other models, in
particular those where the jamming effect is not present or
plays no significant role. These include deposition of
monomer particles, usually of atomic dimensions,
which align with the underlying
lattice without jamming, as well as models where many
layers are formed (mentioned in the preceding section).

The $2D$ deposition with relaxation
of extended objects is of interest in certain experimental
systems where the depositing objects are proteins [17].
Here we focus on the combined effect of jamming and
diffusion, and emphasize dynamics at large times.
For early stages of the deposition process, low-density
approximation schemes can be used. One such application
was reported [34] for continuum deposition of circles
on a plane.

In order to identify new features characteristic of $2D$, 
let us consider
deposition of $2 \times 2$ squares on the square lattice.
The particles are exactly aligned with the $2 \times 2$
lattice sites as shown in Figure~4. Furthermore, we assume
that the diffusional hopping is along the lattice
directions $\pm x$ and $\pm y$, one lattice spacing at a
time. In this model dense configurations involve domains
of four phases as shown in Figure~4. As a result,
immobile fragments of empty area can exist. Each such
single-site vacancy (Figure~4) serves as a meeting point of
four domain walls. 

Here by ``immobile'' we mean that the vacancy cannot
move due to local motion of the surrounding particles. For
it to move, a larger empty-area fragment must first arrive,
along one of the domain walls. One such larger empty void
is shown in Figure~4. Note that it serves as a kink in the
domain wall.
Existence of locally immobile (``frozen'') vacancies suggests possible
frozen glassy behavior with extremely slow relaxation,
at least locally. The full characterization of the
dynamics of this model requires further study. The first
numerical results [77] do provide some answers which
will be reviewed shortly. 

\noindent{\it 4.2. Ordering by Shortening of Domain Walls}

We first consider a
simpler model depicted in Figure~5. In this model [78-79]
the extended particles are squares of size
$\sqrt{2} \times \sqrt{2}$. They are rotated 45$^\circ$
with respect to the underlying square lattice. Their
diffusion, however, is along the vertical and horizontal
lattice axes, by hopping one lattice
spacing at a time. The equilibrium variant of this model
(without deposition, with fixed particle density) is the
well-studied hard-square model [82] which, at large
densities, phase separates into two distinct phases. These
two phases also play role in the late stages of RSA with
diffusion. Indeed, at large densities the 
empty area is stored in domain walls separating
ordered regions. One such domain wall is shown in
Figure~5. Snapshots of actual Monte Carlo simulation
results can be found in Refs.~78-79.

Figure~5 illustrates the process of ordering which
essentially amounts to shortening of domain walls. In
Figure~5, the domain wall gets shorter after the shaded
particles diffusively rearrange to open up a deposition
slot which can be covered by an arriving particle.
Numerical simulations [78-79] find behavior reminiscent of the
low-temperature equilibrium ordering processes [83-85] driven
by diffusive evolution of the domain-wall structure. For
instance, the remaining uncovered area vanishes according
to

$$ 1 - \rho (t) \sim { 1 \over \sqrt{t} } \;\; . \eqno(11) $$

\noindent{}This quantity, however, also measures the length
of domain walls in the system (at large times). Thus,
disregarding finite-size effects and assuming that the
domain walls are not too convoluted (as confirmed by
numerical simulations), we conclude that the power law
Eq.~(11) corresponds to typical domain sizes growing as $\sim
\sqrt{t}$, reminiscent of the equilibrium ordering
processes of systems with nonconserved order parameter
dynamics [83-85].

\noindent{\it 4.3. Numerical Results for Models with Frozen Vacancies}

We now turn again to the $2 \times 2$ model of Figure~4. The
equilibrium variant of this model corresponds to
hard-squares with both nearest and next-nearest neighbor
exclusion [82,86-87]. It has been studied in lesser detail than
the two-phase hard-square model described in the
preceding paragraphs. In fact, the equilibrium phase
transition has not been fully classified (while it was
Ising for the simpler model). The ordering at low
temperatures and high densities was studied [86].
However, many features noted, for instance large entropy of
the ordered arrangements, require further investigation. The
dynamical variant (RSA with diffusion) of this model was
studied numerically [77]. The configuration of the
single-site frozen (locally immobile) vacancies 
and the associated network of
domain walls turn out to be boundary-condition sensitive.
For periodic boundary conditions the density freezes at
values $1-\rho \sim L^{-1}$, where $L$ is the linear system
size.

Preliminary indications were found [77] that the domain
size and shape distributions in such a frozen state are
nontrivial. Extrapolation $L \to \infty$ indicates that
the power law behavior similar to Eq.~(11) is nondiffusive:
the exponent $1/2$ is replaced by $\sim 0.57$. However,
the density of the smallest mobile vacancies, i.e., dimer kinks
in domain walls, one of which is illustrated in
Figure~4, does decrease diffusively. Further studies are
needed to fully clarify the ordering process associated
with the approach to the full coverage as $t \to \infty$ and $L
\to \infty$ in this model.

Even more complicated behaviors are possible when the
depositing objects are not symmetric and can have
several orientations as they reach the substrate. In addition to 
translational diffusion (hopping), one has to consider possible
rotational motion. The square-lattice deposition 
of dimers, with hopping processes including one-lattice-spacing 
motion along the dimer axis and 90$^\circ$ rotations about a
constituent monomer, was studied [80]. The dimers were
allowed to deposit vertically and horizontally. In this 
case, the full close-packed coverage is not
achieved at all because the frozen vacancy sites can be embedded
in, and move by diffusion in, extended structures of
different topologies. These structures are probably less efficiently
demolished by the motion of mobile vacancies than
the elimination of localized frozen vacancies in the model of Figure~4.

\vfil\eject

\noindent{\bf 5.~Conclusion}

In summary, we reviewed theoretical developments in the description
of deposition processes of
extended objects, with jamming and diffusional relaxation. 
While significant progress
has been achieved in $1D$, the $2D$ systems require further
study. Most of these investigations will involve
large-scale numerical simulations.

Other research directions that require further work include multilayer 
deposition and particle detachment, especially the theoretical description
of the latter, including the description of the distribution of 
values/shapes of the primary minimum in the particle-surface interaction potential. This would allow to advance beyond the present theoretical trend of studying deposition as mainly the process of particle transport to the surface, with little or no role played by the 
details of the actual 
particle-surface and particle-particle double-layer and other interactions. 
Ultimately, we would like to interrelate the present deposition studies and approaches in study of adhesion [4], of typically larger particles of sizes up to several microns, at surfaces.

\vfil\eject

\centerline{\bf REFERENCES}

\ 

{\frenchspacing

\item{[1]} {\it Particle Deposition at the Solid-Liquid Interface}, edited by Tardos, Th.F., and Gregory, J., {\it Colloids Surf.}, Vol. 39, No. 1/3, 30 August, 1989.

\item{[2]} {\it Advances in Particle Adhesion}, edited by Rimai, D.S., and 
Sharpe, L.H. (Gordon and Breach Publishers, Amsterdam, 1996).

\item{[3]} {\it Particle Deposition \& Aggregation. Measurement, Modeling and Simulation}, Elimelech, M., Gregory, J., Jia, X., and Williams, R.A. (Butterworth-Heinemann Woburn, MA, 1995).

\item{[4]} {\it Adhesion of Submicron Particles on Solid Surfaces}, edited by Privman, V., {\it Colloids Surf.\/} A (in print, 2000).

\item{[5]} Levich, V.G., {\it Physiochemical Hydrodynamics\/} (Prentice-Hall, London, 1962).

\item{[6]} Privman, V., {\it Trends in Statistical
Physics\/} {\bf 1}, 89 (1994).

\item{[7]} {\it Nonequilibrium Statistical 
Mechanics in One Dimension}, edited by Privman, V.\ (Cambridge 
University Press, 1997).

\item{[8]} Feder, J., and Giaever, I.,
{\it J. Colloid Interface Sci.\/} {\bf 78}, 144 (1980).

\item{[9]} Schmitt, A., Varoqui, R., Uniyal, S., Brash, J.L., and Pusiner, C.,
{\it J. Colloid Interface Sci.\/} {\bf 92}, 25 (1983).

\item{[10]} Onoda, G.Y., and Liniger, E.G.,
{\it Phys. Rev.\/} A{\bf 33}, 715 (1986).

\item{[11]} Kallay, N., Tomi\'c, M., Bi\v skup, B., Kunja\v si\'c,
I., and Matijevi\'c, E.,
{\it Colloids Surf.\/} {\bf 28}, 185 (1987).

\item{[12]} Aptel, J.D., Voegel, J.C., and Schmitt, A.,
{\it Colloids Surf.\/} {\bf 29}, 359 (1988).

\item{[13]} Adamczyk, Z.,
{\it Colloids Surf.\/} {\bf 39}, 1 (1989).

\item{[14]} Adamczyk, Z., Zembala, M., Siwek, B., and Warszy{\' n}ski, P.,
{\it J. Colloid Interface Sci.\/} {\bf 140}, 123 (1990).

\item{[15]} Ryde, N., Kihira, H., and Matijevi\'c, E.,
{\it J. Colloid Interface Sci.\/} {\bf 151}, 421 (1992).

\item{[16]} Song, L., and Elimelech, M.,
{\it Colloids Surf.\/} A{\bf 73}, 49 (1993).

\item{[17]} Ramsden, J.J.,
{\it J. Statist. Phys.\/} {\bf 73}, 853 (1993).

\item{[18]} Murphy, C.J., Arkin, M.R., Jenkins,
Y., Ghatlia, N.D., Bossmann, S.H., Turro, N.J., and Barton, J.K., 
{\it Science\/} {\bf 262}, 1025 (1993).

\item{[19]} Bartelt, M.C., and Privman, V.,
{\it Internat. J. Mod. Phys.\/} B{\bf 5}, 2883 (1991).

\item{[20]} Evans, J.W.,
{\it Rev. Mod. Phys.\/} {\bf 65}, 1281 (1993).

\item{[21]} Widom, B., {\it J. Chem. Phys.\/} {\bf 58}, 4043 (1973).

\item{[22]} Schaaf, P., and Talbot, J.,
{\it Phys. Rev. Lett.\/} {\bf 62}, 175 (1989).

\item{[23]} Dickman, R., Wang, J.-S., and Jensen, I., 
{\it J. Chem. Phys.\/} {\bf 94}, 8252 (1991).

\item{[24]} Gonzalez, J.J., Hemmer, P.C., and H{\o}ye, J.S.,
{\it Chem. Phys.\/} {\bf 3}, 228 (1974).

\item{[25]} Feder, J., {\it J. Theor. Biology\/} {\bf 87}, 237 (1980).

\item{[26]} Tory, E.M., Jodrey, W.S., and Pickard, D.K.,
{\it J. Theor. Biology\/} {\bf 102}, 439 (1983).

\item{[27]} Hinrichsen, E.L., Feder, J., and J\o ssang, T., 
{\it J. Statist. Phys.} {\bf 44}, 793 (1986). 

\item{[28]} Burgos, E., and Bonadeo, H., {\it J. Phys.\/} A{\bf 20}, 1193 (1987).

\item{[29]} Barker, G.C., and Grimson, M.J.,
{\it J. Phys.\/} A{\bf 20}, 2225 (1987).

\item{[30]} Vigil, R.D., and Ziff, R.M.,
{\it J. Chem. Phys.\/} {\bf 91}, 2599 (1989).

\item{[31]} Talbot, J., Tarjus, G., and Schaaf, P.,
{\it Phys. Rev.\/} A{\bf 40}, 4808 (1989).

\item{[32]} Vigil, R.D., and Ziff, R.M.,
{\it J. Chem. Phys.\/} {\bf 93}, 8270 (1990).

\item{[33]} Sherwood, J.D., {\it J. Phys.\/} A{\bf 23}, 2827 (1990).

\item{[34]} Tarjus, G., Schaaf, P., and  Talbot, J.,
{\it J. Chem. Phys.} {\bf 93}, 8352 (1990).

\item{[35]} Brosilow, B.J., Ziff, R.M., and Vigil, R.D.,
{\it Phys. Rev.\/} A{\bf 43}, 631 (1991).

\item{[36]} Privman, V., Wang, J.-S., and Nielaba, P.,
{\it Phys. Rev.} B{\bf 43}, 3366 (1991).

\item{[37]} Pomeau, Y.,
{\it J. Phys.\/} A{\bf 13}, L193 (1980).

\item{[38]} Swendsen, R.H.,
{\it Phys. Rev.\/} A{\bf 24}, 504 (1981).

\item{[39]} Kallay, N., Bi\v skup, B., Tomi\'c, M., and Matijevi\'c, E.,
{\it J. Colloid Interface Sci.\/} {\bf 114}, 357 (1986).

\item{[40]} Barma, M., Grynberg, M.D., and Stinchcombe, R.B.,
{\it Phys. Rev. Lett.\/} {\bf 70}, 1033 (1993).

\item{[41]} Stinchcombe, R.B., Grynberg, M.D., and Barma, M.,
{\it Phys. Rev.\/} E{\bf 47}, 4018 (1993).

\item{[42]} Grynberg, M.D., Newman, T.J., and Stinchcombe, R.B.,
{\it Phys. Rev.\/} E{\bf 50}, 957 (1994).

\item{[43]} Grynberg, M.D., and Stinchcombe, R.B.,
{\it Phys. Rev.\/} E{\bf 49}, R23 (1994).

\item{[44]} Sch\"utz, G.M., {\it J. Statist. Phys.\/} {\bf 79}, 243 (1995).

\item{[45]} Krapivsky, P.L., and Ben-Naim, E.,
{\it J. Chem. Phys.\/} {\bf 100}, 6778 (1994). 

\item{[46]} Barma, M., and Dhar, D.,
{\it Phys. Rev. Lett.\/} {\bf 73}, 2135 (1994).

\item{[47]} Bossmann, S.H., and Schulman, L.S.,
in {\it Nonequilibrium Statistical 
Mechanics in One Dimension}, edited by Privman, V. (Cambridge 
University Press, 1997), p. 443.

\item{[48]} Privman, V., and Barma, M.,
{\it J. Chem. Phys.\/} {\bf 97}, 6714 (1992).

\item{[49} Nielaba, P., and Privman, V.,
{\it Mod. Phys. Lett.\/} B {\bf 6}, 533 (1992).

\item{[50]} Bonnier, B., and McCabe, J., 
{\it Europhys. Lett.\/} {\bf 25}, 399 (1994).

\item{[51]} Kang, K., Meakin, P., Oh, J.H., and Redner, S.,
{\it J. Phys.\/} A {\bf 17}, L665 (1984).

\item{[52]} Cornell, S., Droz, M., and Chopard, B.,
{\it Phys. Rev.\/} A{\bf 44}, 4826 (1991).

\item{[53]} Privman, V., and Grynberg, M.D.,
{\it J. Phys. A\/} {\bf 25}, 6575 (1992).

\item{[54]} ben-Avraham, D.,
{\it Phys. Rev. Lett.\/} {\bf 71}, 3733 (1993).

\item{[55]} Krapivsky, P.L.,
{\it Phys. Rev.\/} E {\bf 49}, 3223 (1994).

\item{[56]} Lee, B.P.,
{\it J. Phys.\/} A {\bf 27}, 2533 (1994).

\item{[57]} Grynberg, M.D.,
{\it Phys. Rev.\/} E {\bf 57}, 74 (1998).

\item{[58]} Kang, K., and Redner, S.,
{\it Phys. Rev. Lett.\/} {\bf 52}, 955 (1984).

\item{[59]} Kang, K., and Redner, S.,
{\it Phys. Rev.\/} A{\bf 32}, 435 (1985).

\item{[60]} Racz, Z.,
{\it Phys. Rev. Lett.\/} {\bf 55}, 1707 (1985).

\item{[61]} Bramson, M., and Lebowitz, J.L.,
{\it Phys. Rev. Lett.\/} {\bf 61}, 2397 (1988).

\item{[62]} Balding, D.J., and Green, N.J.B.,
{\it Phys. Rev.\/} A {\bf 40}, 4585 (1989).

\item{[63]} Amar, J.G., and Family, F.,
{\it Phys. Rev.\/} A {\bf 41}, 3258 (1990).

\item{[64]} ben-Avraham, D., Burschka, M.A., and Doering, C.R.,
{\it J. Statist. Phys.\/} {\bf 60}, 695 (1990).

\item{[65]} Bramson, M., and Lebowitz, J.L.,
{\it J. Statist. Phys.\/} {\bf 62}, 297 (1991).

\item{[66]} Privman, V.,
{\it J. Statist. Phys.\/} {\bf 69}, 629 (1992).

\item{[67]} Privman, V., and Nielaba, P.,
{\it Europhys. Lett.\/} {\bf 18}, 673 (1992).

\item{[68]} Grynberg, M.D., and Stinchcombe,  R.B.,
{\it Phys. Rev. Lett.\/} {\bf 74}, 1242 (1995).

\item{[69]} Gan, C.K., and Wang, J.-S.,
{\it Phys. Rev.\/} E{\bf 55}, 107 (1997).

\item{[70]} de Oliveira, M.J., and Petri, A.,
{\it J. Phys.\/} A{\bf 31}, L425 (1998).

\item{[71]} Xiao, R.-F., Alexander, J.I.D., and Rosenberger, F.,
{\it Phys. Rev.\/} A{\bf 45}, R571 (1992).

\item{[72]} Lubachevsky, B.D., Privman, V., and Roy, S.C.,
{\it Phys. Rev.\/} E{\bf 47}, 48 (1993).

\item{[73]} Lubachevsky, B.D., Privman, V., and Roy, S.C.,
{\it J. Comp. Phys.\/} {\bf 126}, 152 (1996).

\item{[74]} Privman, V., Frisch, H.L., Ryde, N., and Matijevi\'c, E.,
{\it J. Chem. Soc. Farad. Tran.\/} {\bf 87}, 1371 (1991).

\item{[75]} Ryde, N., Kallay, N., and
Matijevi\'c, E.,
{\it J. Chem. Soc. Farad. Tran.\/} {\bf 87}, 1377 (1991).

\item{[76]} Zelenev, A., Privman, V., and Matijevi\'c, E., 
{\it Colloids Surf.\/} A{\bf 135}, 1 (1998).

\item{[77]} Wang, J.-S., Nielaba, P., and Privman, V.,
{\it Physica\/} A{\bf 199}, 527 (1993).

\item{[78]} Wang, J.-S., Nielaba, P., and Privman, V.,
{\it Mod. Phys. Lett.\/} B{\bf 7}, 189 (1993).

\item{[79]} James, E.W., Liu, D.-J., and Evans, J.W., in Ref.~4.

\item{[80]} Grigera, S.A., Grigera, T.S., and Grigera, J.R.,
{\it Phys. Lett\/} A{\bf 226}, 124 (1997).

\item{[81]} Vernables, J.A., Spiller, G.D.T., and Hanb\"ucken, M.,
{\it Rept. Prog. Phys.\/} {\bf 47}, 399 (1984).

\item{[82]} Runnels, L.K., in {\it Phase Transitions and Critical
Phenomena}, Vol. 2, edited by Domb, C., and Green, M.S.
(Academic, London, 1972), p. 305.

\item{[83]} Gunton, J.D., San Miguel, M., and Sahni, P.S., in
{\it Phase Transitions and Critical
Phenomena}, Vol. 8, edited by Domb, C., and Lebowitz, J.L.
(Academic, London, 1983), p. 267.

\item{[84]} Mouritsen, O.G., in {\it Kinetics of
Ordering and Growth at Surfaces}, edited by Lagally, M.G.
(Plenum, NY, 1990), p. 1.

\item{[85]} Sadiq, A., and Binder, K., {\it J. Statist. Phys.\/} {\bf 35}, 517
(1984).

\item{[86]} Binder, K., and Landau, D.P.,
{\it Phys. Rev.\/} B{\bf 21}, 1941 (1980).

\item{[87]} Kinzel, W., and Schick, M.,
{\it Phys. Rev.\/} B{\bf 24}, 324 (1981).

}

\vfil\eject

\noindent{\bf Figure Captions}

\noindent\hang{}Figure 1: Possible configurations of particles at surfaces. From left to right, $A$ --- particles 
deposited directly on the collector; $B$ --- particles deposited on top of other particles. We next show an example of jamming, $C$ --- a particle marked by an open circle cannot fit in the lowest layer at the surface. A top view of a more realistic two-dimensional ($2D$) surface configuration is shown in the inset. The rightmost example, $E$, illustrates screening:
surface position marked by the open circle is not reachable.

\noindent\hang{}Figure 2: Deposition of dimers on the $1D$
lattice. Only one of the three hatched dimers can deposit on the
surface, which then becomes fully jammed in the interval shown.

\noindent\hang{}Figure 3: Schematic variation of the
coverage $\rho (t)$ with time for
deposition without
(lower curve) and with (upper curve) diffusional or
other relaxation. The ``ordered'' density corresponds to
close packing. 

\noindent\hang{}Figure 4: Fragment of a 
deposit configuration in the deposition of $2\times 2$
squares. Illustrated are one single-site frozen vacancy
at which four domain walls meet (indicated by arrows),
and one dimer
vacancy which causes a kink in one of the domain walls.

\noindent\hang{}Figure 5: Illustration of deposition of
$\sqrt{2} \times \sqrt{2}$ particles on the square lattice. 
Diffusional motion
during time interval from $t_1$ to $t_2$ can rearrange the
empty area ``stored'' in the domain wall to open up a new
landing site for deposition. This is illustrated by the
shaded particles.

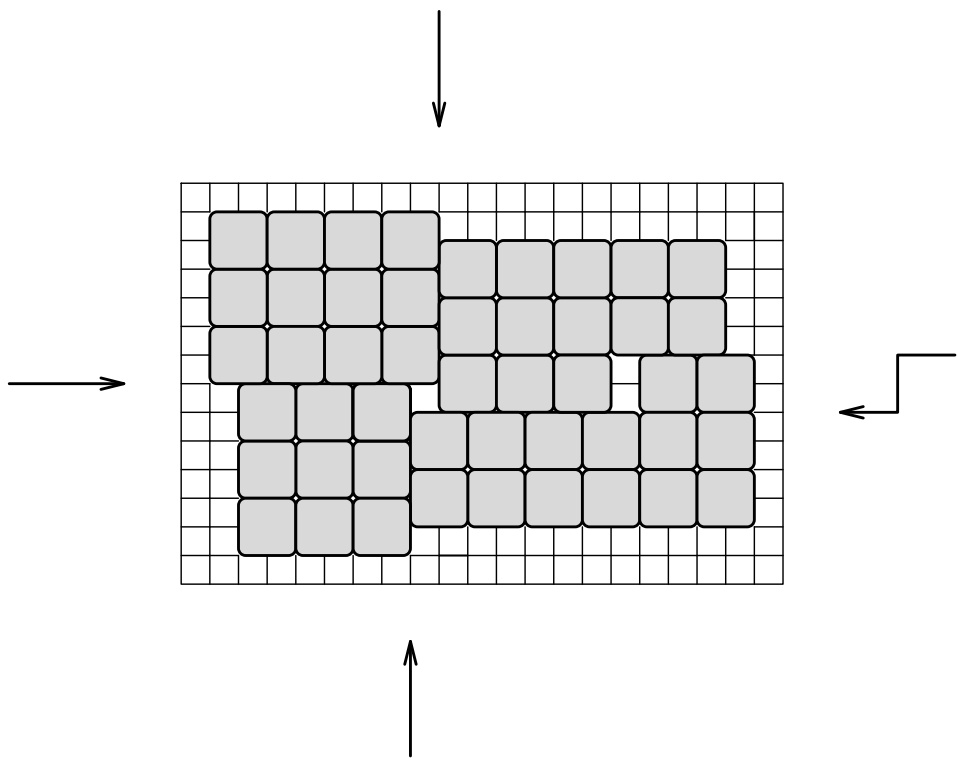
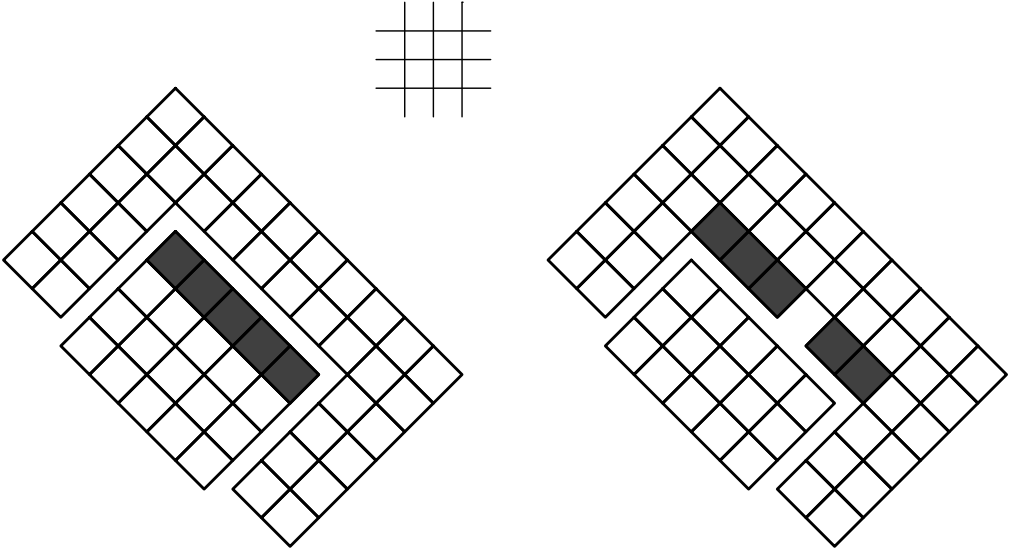
\bye